\documentstyle[12pt,epsf]{article}
\def\DRAFT{0}
\renewcommand{\theequation}{\thesection.\arabic{equation}}

\newcommand{\eq}{\begin{equation}}
\newcommand{\en}{\end{equation}}

\newcommand{\eqa}{\begin{eqnarray}}
\newcommand{\ena}{\end{eqnarray}}
\newcommand{\eqan}{\begin{eqnarray*}}
\newcommand{\enan}{\end{eqnarray*}}
\newcommand{\bear}[1]{\begin{array}{#1}}
\newcommand{\enar}{\end{array}}

\newcommand{\lbl}[1]{ \ifnum \DRAFT = 0
                             {\label {#1}}
                      \else  {\makebox[0in]{\raisebox{-2ex}{\tiny #1}
                                            \hspace{-6ex}}}
                             {\label {#1}}
                      \fi}
\newcommand{\rf}[1]{  \ifnum \DRAFT = 0
                             \ref {#1}
                      \else {\ref {#1}}
                            \makebox[0in]{\raisebox{-2ex}{\tiny #1}}
                      \fi}
\newcommand{\ct}[1]{  \ifnum \DRAFT = 0
                             \cite {#1}
                      \else {\cite {#1}}
                            \makebox[0in]{\raisebox{-2ex}{\tiny #1}}
                      \fi}

\newlength{\superigor}
\newcommand{\spazio}[1]{\settowidth{\superigor}{${#1}$}
                        \makebox[\superigor]{} }
\def \eroinappendice{0}
\newcounter{temporaneo}
\newcounter{appendice}

\newcommand{\sect}[1]{
\ifnum \eroinappendice =1
       \def\eroinappendice{0}
       \setcounter{appendice}{\value{section}}
       \setcounter{section}{\value{temporaneo}}
       \renewcommand{\theequation}{\thesection.\arabic{equation}}
       \fi
\setcounter{equation}{0}\section{#1}}

\newcommand{\app}[1]{
\ifnum \eroinappendice =0
       \def\eroinappendice{1}
       \setcounter{temporaneo}{\value{section}}
       \setcounter{section}{\value{appendice}}
       \renewcommand{\thesection}{Appendix \Alph{section} }
       \renewcommand{\theequation}{\Alph{section}.\arabic{equation}}
      \fi
\setcounter{equation}{0}\section{#1}}

\newcommand{\draft}{ \ifnum \DRAFT =0  {} \else {\\ DRAFT} \fi}

\newcommand{\IJMP}[1]{Int. Jou. Mod. Phys. \ {\bf #1}\ }

\newcommand{\MPL}[1]{Mod. Phys. Lett.\ {\bf #1}\ }

\newcommand{\PR}[1]{Phys. Rev\ {\bf #1}\ }

\newcommand{\PTP}[1]{Prog. Theor. Phys.\ {\bf #1}\ }

\def\sqr#1#2{{\vcenter{\hrule height.#2pt
     \hbox{\vrule width.#2pt height#1pt \kern#1pt
        \vrule width.#2pt}
     \hrule height.#2pt}}}

\def\thinspace{\kern .16667em}

\def\Dir{\nabla\kern-2ex\Big{/}}
\def\dslash{\partial\kern-1.5ex\Big{/}}

\def\reali{{\hbox{\s@ l\kern-.5ex R}}}
\def\naturali{{\hbox{\s@ l\kern-.5ex N}}}
\def\interi{{\mathchoice
 {\hbox{Z\kern-1.5mm Z}}
 {\hbox{Z\kern-1.5mm Z}}
 {\hbox{{Z\kern-1.2mm Z}}}
 {\hbox{{Z\kern-1.2mm Z}}}  }}

\def\unity{{\hbox{\s@ 1\kern-.8mm l}}}
\def\uno{{\hbox{ 1\kern-.8mm l}}}
\def\part{\partial}

\def\rd{\sqrt{2}}
\def\um{{1\over2}}
\def\usrd{{1\over\sqrt{2}}}

\def\Llrarr{\Longleftrightarrow}

\def\lrarr{\leftrightarrow}
\def\rarr{\rightarrow}
\def\larr{\leftarrow}

\def\ot{\otimes}

\def\dag{\dagger}

\def\CA{{\cal A}}
\def\CC{{\cal C}}

\def\CI{{\cal I}}
\def\CL{{\cal L}}
\def\CM{{\cal M}}
\def\CN{{\cal N}}

\def\aa{\alpha}
\def\bb{\beta}
\def\cc{\chi}
\def\cb{\bar\chi}
\def\dd{\delta}
\def\DD{\Delta}
\def\ee{\epsilon}

\def\ff{\phi}

\def\FF{\Phi}

\def\gg{\gamma}

\def\LL{\Lambda}

\def\pp{\psi}

\def\pb{\bar\psi}

\def\SS{\Sigma}

\newcommand{\mat}[4]{\left(
                     \begin{array}{cc}
                     {#1} & {#2} \\
                     {#3} & {#4}
                     \end{array}
                     \right)
                    }
\newcommand{\vett}[2]{\left(
                      \begin{array}{c}
                     {#1} \\
                     {#2}
                     \end{array}
                     \right)
                    }
\newcommand{\vet}[2]{\left(
                     \begin{array}{cc}
                     {#1} &  {#2}
                     \end{array}
                     \right)
                    }

\begin{document}

\begin{titlepage}

\begin{flushright}
NORDITA-95/42 P\\
May 1995\\
hep-th/yymmddd
\end{flushright}
\vspace*{0.5cm}

\begin{center}
{\bf
\begin{Large}
{\bf
THE GENERALIZED GROSS-NEVEU MODEL ON THE LIGHT CONE
\\}
\end{Large}
}
\vspace*{1.5cm}
          {\large Igor Pesando}\footnote{E-mail PESANDO@NBIVAX.NBI.DK,
            pesando@hetws4.nbi.dk, 31890::I\_PESANDO}
         \\[.3cm]
          NORDITA\\
          Blegdamsvej 17, DK-2100 Copenhagen \O \\
          Denmark\\
\end{center}
\vspace*{0.7cm}
\begin{abstract}
{
We investigate the generalized Gross-Neveu model
using the discretized light cone quantization
and we find that the vacuum of the bare theory is {\sl non}
trivial in presence of vectorial current coupling when the simplest
and most natural form of quantum constraints is used. Nevertheless the
vacuum of the renormalized theory is trivial.

In the thermodynamic the Bethe-Salpiter equations which are obtained
contain all the terms needed to make them finite.
}
\end{abstract}
\vfill
\end{titlepage}

\setcounter{footnote}{0}

\def\gs{{g_s}}
\def\gp{{g_p}}
\def\gv{{g_v}}
\def\ub{{\underline{b}}}
\def\ur{{\underline{r}}}
\def\us{{\underline{s}}}
\def\ut{{\underline{t}}}
\def\ovr{{\overline{r}}}
\def\os{{\overline{s}}}
\def\ot{{\overline{t}}}
\def\BS{Bethe-Salpiter~}

\sect{Introduction}
The generalized Gross-Neveu model (\ct{GN}) has been a subject of research in
the last two decades because of the many interesting features such as
dynamical symmetry breaking, asymptotic freedom etc.
The generalized Gross-Neveu model was already studied on the light
cone in ref. (\ct{TO}) where the \BS equations were obtained in an indirect
way without explicitely solving the constraints which arise from the
equations of motion of the non propagating fields and in the usual
noncompact Minkowsky space.
In so doing the authors
missed some interesting features like the fact that the hamiltonian
vanishes in the massless case, the explicit appearance of the running
coupling constants and the finitness of the \BS equations in the
thermodynamical limit.

In a previous paper (\ct{Pe}) we examined the pure (massive) Gross-Neveu
model using the discretized light cone quantization (DCLQ), i.e.
on the light cone cylinder and we were able to find a nice expression
for the hamiltonian $P^-$ to all orders in $1\over N$ involving
only the running coupling constant and a finite \BS equation.
Here we examime the generalized Gross-Neveu model in order to see
whether all these nice features, we found in the Gross-Neveu model,
survive in a different model.
It turns out that we cannot give a nice
explicit expression for $P^-$ to all orders but the other and
most interesting propriety, i.e. the finiteness of the \BS equation in
the thermodynamical limit is maintained.
This feature seems to be quite universal.
Also in the case of $QED_{1+3}$ (\ct{KR}) it has been in fact
observed  that in the thermodynamical limit terms which improve
the UV behaviour are generated.
We find also an unexpected new feature: the appearance of a non
trivial vacuum in the bare (regularized) action.

\sect{The generalized Gross-Neveu model on the light cone.}
The lagrangian of the (massive) generalized  Gross-Neveu model
(\cite{GN}) is given by (notice that $\gs$ has the opposite sign
w.r.t. the usual one, in particular $\gs=-g^2$ w.r.t. the notation of
ref. (\ct{Pe}) )
\eq
\CL=\pb\cdot({i} \stackrel{\lrarr}{\dslash}-m)\pp
-{\gs\over N}(\pb\cdot\pp)^2
-{\gp\over N}(\pb\cdot\gg_5\pp)^2
-{\gv\over N}(\pb\cdot\gg_\mu\pp)^2
\en
that can be explicitly written in the light cone as\footnote{
{\bf Conventions.}
$$
x^\pm=x_\mp=\usrd(x^0\pm x^1)~~~~
A^\mu B_\mu=
A_0 B_0 - A_1 B_1=
A_+ B_- + A_+ B_-
{}~~~~
\ee^{01}=-\ee^{+-}=1
$$
$$
\gg_+=\mat{0}{\rd}{0}{0}~~~~
\gg_-=\mat{0}{0}{\rd}{0}~~~~
\gg_0=\mat{0}{1}{1}{0}~~~~
\gg_1=\mat{0}{1}{-1}{0}~~~~
$$
$$
\gg_5=-\gg_0\gg_1=\mat{1}{0}{0}{-1}~~~~
P_{R,L}={1\pm\gg_5\over2}
$$
$$
\pp=\vett{\pp}{\cc}~~~~
\pb=\vet{\cb}{\pb}~~~~
 \cc\pb=-\usrd
\mat
{\rd\pb P_R\cc}
{\pb \gg_-\cc}
{\pb \gg_+\cc}
{\rd\pb P_L\cc}
$$
$$
\int_x=\int d^2 x ~~~~
\int_p=\int {d^2 p\over (2\pi)^2}
$$
$$
\stackrel{\lrarr}{\part}=\um(\stackrel{\rarr}{\part}-\stackrel{\larr}{\part})
$$
$$
r,s,t\in \interi+\um ~~~~
m,n\in \interi~~~~
$$
}

\eqa
\CL&=&i{\rd}(\pb\cdot\stackrel{\lrarr}{\part_+}\pp
+\cb\cdot\stackrel{\lrarr}{\part_-}\cc)
-m(\pb\cdot\cc +\cb\cdot\pp)
\nonumber\\
&&
-{\gs+\gp\over N}\left[ (\pb\cdot\cc)^2 + (\cb\cdot\pp)^2 \right]
-2{\gs-\gp\over N}\pb\cdot\cc~\cb\cdot\pp
-4{\gv\over N}\pb\cdot\pp \cb\cdot\cc
\lbl{gn-lag}
\nonumber\\~
\ena
where $\pp=(\pp^i)=\pb^*$ with $i=1\dots N$.
As it is usual in the light cone approach primary constraints are
given by the classical equation of motion for the nonpropagating fields
$\cb^i$
\eq
i\rd \part_-\cc^i -m\pp^i
-2{\gs+\gp\over N}\pp^i~\cb\cdot\pp
-2{\gs-\gp\over N}\pp^i~\pb\cdot\cc
-4{\gv\over N}\cc^i~\pb\cdot\pp
=0
\lbl{eq-mot-b}
\en
and $\cc^i$
\eq
-i\rd \part_-\cb^i -m\pb^i
-2{\gs+\gp\over N}\pb\cdot\cc~\pb^i
-2{\gs-\gp\over N}\cb\cdot\pp~\pb^i
-4{\gv\over N}\pb\cdot\pp~\cb^i
=0
\lbl{eq-mot}
\en
Using these constraints we can rewrite the lagrangian (\rf{gn-lag}) as
\eq
\CL'=i\rd\pb\cdot\part_+\pp
-{m\over2}(\cb\cdot\pp +\pb\cdot\cc)
\en
where $\cc$ is to be seen as a functional of $\pp$.
{}From the previous effective lagrangian we get the translation generators
\eqa
{ P}^-
&=&{m\over 2}\int d x^-
\nonumber\\
{ P}^+
&=&i\rd\int dx^- \pb\cdot\part_-\pp
\lbl{tran-gen}
\ena
Notice that when $m=0$ $P^-$ vanishes exactly as in the pure
Gross-Neveu model (\ct{Pe})
and hence we cannot quantize the massless model on the light cone.
These generators are hermitian because we started from a real lagrangian.
In particular in order to have a real lagrangian we have written the kinetic
term as $(\pb\cdot\stackrel{\lrarr}{\dslash}\pb)$.

We quantize the theory imposing  the standard Dirac brackets
\eq
\{\pp^i(x),\pb^j(y)\}|_{x^+=y^+}=\usrd \dd^{i j} \dd(x^--y^-)
\lbl{can-com-rel}
\en
in the light cone box $x^-\in [-L, L]$ with  the standard
antiperiodic boundary condition
\eq
\pp^i(x^- +2L)=-\pp^i(x^-)
{}~~~~
\pb^i(x^- +2L)=-\pb^i(x^-)
\en
Expanding the operator $\pp_+$ in Schr\"odinger picture in Fourier
modes
\eqa
\pp^i(x)={1\over \sqrt[4]{2}}\sum_{r\in\interi+\um}\pp_r^i
                             {e^{\pi i r {x\over L}}\over\sqrt{2L}}
\nonumber\\
\pb^i(x)={1\over \sqrt[4]{2}}\sum_{r\in\interi+\um}\pb_r^i
                             {e^{-\pi i r {x\over L}}\over\sqrt{2L}}
\ena
we see that the anticommutation relations in eq. (\rf{can-com-rel})
imply:
\eq
\{\pp_r^i,\pb_s^j\}=\dd_{r s}\dd^{i j}
\en
With in mind the idea of using a variational approach
to find the state minimizing the energy (which is
defined to be the eigenvalue of $P^-$), we introduce the normal order
$N_\CA[\dots]$ defined by
\eq
\pp_r=
\left\{
\bear{l}
 {\mbox{if }~r\in\CC~\mbox{creation operator}}
\\
 {\mbox{if }~r\in\CA~\mbox{annihilation operator}}
\enar
\right.
{}~~~~
\pb_r=
\left\{
\bear{l}
 {\mbox{if }~r\in{\bar\CA}=\CC~\mbox{annihilation operator}}
\\
 {\mbox{if }~r\in{\bar\CC}=\CA~\mbox{creation operator}}
\enar
\right.
\en
where $\CA\cup\CC=\interi+\um$, $\CA\cap\CC=\oslash$
\footnote{
With the trivial perturbative vacuum, which is defined by
$\CA=\{ {1\over2},{3\over2},{5\over2},\dots\}$ and
$\CC=\{ -{1\over2},-{3\over2},-{5\over2},\dots\}$, and
in the usual notation we would have
$$\pp_r=\left\{
\bear{ll}
d^\dag_{-r} &{\mbox{if }~r\in\CC}
\\
b_r         &{\mbox{if }~r\in\CA}
\enar
\right.
{}~~~~
\pb_r=
\left\{
\bear{ll}
d_{-r}   &{\mbox{if }~r\in{\bar\CA}=\CC}
\\
b_r^\dag &{\mbox{if }~r\in{\bar\CC}=\CA}
\enar
\right.
$$
}.
Since the action is C-invariant, we require that the vacuum $|\CA>$ be
C-invariant\footnote{
We define the charge conjugation as $C\pp(x^-)C^{-1}=\pb(x^-) \Llrarr
C\pp_rC^{-1}=\pb_{-r}$.
}
and we have to impose $r\in\CA\Llrarr -r\in\CC$.
The choice of the set $\CA$ is equivalent to consider as vacuum the state
\eq
|\CA>\propto
{\prod_{s\in\CA}\prod_{i=1}^N \pp^i_s|0>}
\en
where $|0>$ is the usual free vacuum, defined as $\pp_{-r}|0>=\pb_r|0>=0$ for
$r>0$.

After this introductory stuff we can try to solve the constraints explicitly
and then to write down the explicit form of the translation generators
(\rf{tran-gen}).
Differently from the pure Gross-Neveu model (\ct{Pe})
we are obliged to solve the constraints explicitely and this requires
a slightly different technique.

We discuss the logic of the computation in appendix A and we give some
intermediate results in appendix B.

The explicit and lengthy computation yields
\def\ju{{J\kern-0.0em\makebox[0in]{\raisebox{-1ex}{\tiny 1}}\kern0.5em}}
\def\jd{{J\kern-0.0em\makebox[0in]{\raisebox{-1ex}{\tiny 2}}\kern0.5em}}
\def\jz{{J\kern-0.0em\makebox[0in]{\raisebox{-1ex}{\tiny 0}}\kern0.5em}}

\eq
P^+= -N {\pi\over L} \sum_{r\in\CC} r
-{\pi\over L} \sum_r r N[\pb_r\cdot\pp_r]
\en
and
\eqa
P^-&=&
N{m^2 L\over 2\pi} {\SS(0)\over 1-{\gs\over\pi}\SS(0)}
-{M^2 L\over 2\pi}
\left(\sum_r {N[\pb_r\cdot\pp_r]\over r+\aa}
    -{\gv\over\pi}\sum_r N[\pb_r\cdot\pp_r] \sum_t {\DD_t\over(t-\aa)^2}
\right)
\nonumber\\
&&+{1\over N}{M^2 L\over 2\pi}
\Biggr[ \sum_{n}
        \sum_p {N[\pb_p\cdot\pp_{p-n}]\over p+\aa}
        \sum_q {N[\pb_{q-n}\cdot\pp_{q}]\over q+\aa}
        \ju_n
\nonumber\\
&&      + \sum_{n}
        \sum_p {N[\pb_p\cdot\pp_{p-n}]\over p+\aa}
        \sum_q {N[\pb_{q}\cdot\pp_{q+n}]\over q+\aa}
        \jd_n
\nonumber\\
&&      + \sum_{n}
        \sum_p {N[\pb_{p-n}\cdot\pp_{p}]\over p+\aa}
        \sum_q {N[\pb_{q+n}\cdot\pp_{q}]\over q+\aa}
        \jd_n
\nonumber\\
&&      +{\gv\over\pi}\sum_{n} \sum_p N[\pb_p\cdot\pp_{p-n}]
        \sum_q {N[\pb_{q}\cdot\pp_{q+n}]\over (q+\aa)(q+\aa+n)}
\nonumber\\
&&      +{\gv\over\pi}\sum_{n} \sum_p N[\pb_p\cdot\pp_{p+n}]
         \sum_q {N[\pb_{q-n}\cdot\pp_{q}]\over q+\aa}
         \jz_n
\nonumber\\
&&      +{\gv\over\pi}\sum_{n} \sum_p N[\pb_p\cdot\pp_{p-n}]
         \sum_q {N[\pb_{q}\cdot\pp_{q-n}]\over q+\aa}
         \jz_n
\nonumber\\
&&      +\left({\gv\over\pi}\right)^2
               \sum_{n} \sum_p N[\pb_p\cdot\pp_{p-n}]
               \sum_q {N[\pb_{q}\cdot\pp_{q+n}]}
\nonumber\\
&&\spazio{+\left({\gv\over\pi}\right)^2\sum_{n}}
  \left( \sum_t {\DD_t\over (t-\aa)^2 (t-\aa-n)}
        +\jz_n \sum_t {\DD_t\over (t-\aa) (t-\aa-n)}
  \right)
\Biggr]
\nonumber\\
\lbl{P-}
\ena
where we have used the symbols $\SS(n), M^2, \ju, \jd, \jz$ defined as
\eqa
\SS(n)&=&\sum_t {\DD_t\over t-\aa+n}
{}~~~~\DD_t=\left\{\bear{ll} 1 & t\in\CA \\ 0 & t\in\CC \enar \right.
\nonumber\\
M^2&=& {m^2\over (1- {\gs\over\pi} \SS(0) )^2}
\nonumber\\
\ju_n&=&{  {\gs-\gp\over 2\pi} + {\gs\over\pi} {\gp\over\pi} \SS(n)
    \over 1 - {\gs-\gp\over 2\pi} (\SS(n)+\SS(-n))
            - {\gs\over\pi} {\gp\over\pi}\SS(n)\SS(-n) }
\nonumber\\
\jd_n&=&{  {\gs+\gp\over 2\pi}
    \over 1 - {\gs-\gp\over 2\pi} (\SS(n)+\SS(-n))
            - {\gs\over\pi} {\gp\over\pi}\SS(n)\SS(-n) }
\nonumber\\
\jz_n&=&-\ju_n \sum_t {\DD_t\over (t-\aa) (t-\aa-n)}
        -\jd_n \sum_t {\DD_t\over (t-\aa) (t-\aa+n)}
\lbl{def0}
\ena

\sect{The simplest case of the pure vector currents interaction}

In order to understand how to treat the previous expression in its
generality and to understand the meaning of the shift
$\aa={\gv\over\pi}(\LL+\um)$
in the previous formulae eq.s (\rf{def0}) and
in the inverse derivative $D^{-1}$ (\rf{D}) we set
$
\gs=\gp=0
$.
In this case we find $\ju=\jd=\jz=0$ and the expression for $P^-$
eq. (\rf{P-}) simplifies a lot; in particular the vacuum energy
becomes simply
\eq
P^-_{\mbox{vacuum}}=
N{m^2 L\over 2\pi}\sum_t {\DD_t\over t-\aa}
=N{m^2 L\over 2\pi}\sum_{t\in\CA} {1\over t-\aa}
\en
It is immediate to realize that the state of minimum energy among the
test states, i.e. the vacuum is given by
\eqa
r\in\CA &\Llrarr& -\LL\le r< -|\aa|~\&~ 0<r<|\aa|~~~~
 \left|{\gv\over\pi}\right|<1
\nonumber\\
r\in\CA &\Llrarr& 0<r\le \LL~~~~
 \left|{\gv\over\pi}\right|>1
\lbl{vacuum}
\ena
This seems odd; we would like to preserve the trivial structure
of the vacuum but the only way to get the trivial vacuum is to require
$|\aa|<\um$ but this implies that
$\left|{\gv\over\pi}\right|<{1\over 2\LL+1}$
and this has the unfortunate consequence of the complete
decoupling of $\gv$ in the \BS-'t Hooft equation when taking the limit
$\LL\rarr\infty$.
But this is not correct.
Hence we do not assume $|\aa|<\um$ and therefore the bare theory has a
non trivial vacuum (\rf{vacuum}) with $|\aa|=O(\LL)$.

In the following we assume $\left|{\gv\over\pi}\right|<1$ since
otherwise there are  discontinuities in $P^-_{\mbox{vacuum}}$ in the
limit $\LL \rarr \infty$
($\lim_{\LL \rarr \infty} P^-_{\mbox{vacuum}}|_{\left|{\gv\over\pi}\right|<1}
 = -\infty$ while
$\lim_{\LL \rarr \infty} P^-_{\mbox{vacuum}}|_{\left|{\gv\over\pi}\right|>1}
=\mbox{finite}$)
and it does not seem possible to find a
sensible theory for $\left|{\gv\over\pi}\right|>1$

We introduce for convenience the shifted indices $\ovr=r+\aa$ which
vary in the range
\eq
\ovr\in\CA \Llrarr -\LL+\aa\le \ovr < -|\aa|+\aa~\&~ \aa<\ovr<|\aa|+\aa
\en
When we take the limit $\LL\rarr\infty$ one of
the two intervals (which of the two depends on the sign of $\gv$)
gives  vanishing small contributions to all the expressions and hence
decouples while the other simulates the usual trivial vacuum.
This can be easily seen in the study of the \BS equation.
Let us define the following normalized ``mesonic'' state
\eq
|\ff,R>={1\over \sqrt N}\sum_{r\in\CC,r-R\in\CA} \pb_{r-R}\cdot\pp_r~~
\ff_R(r) |0>
\lbl{meson}
\en
which has momentum $\pi R\over L$, i.e.
$P^+|\ff,R>={\pi R\over L}|\ff,R>$ and wave function
\eq
\ff_R(r)=\FF_R(\ovr)
\en
The \BS equation then reads
\eqa
M^2_{\mbox{meson}}\FF_R(\os)&=& m^2 {R^2\over \os(R-\os)} \FF_R(\os)
-m^2 {\gv\over\pi} {R\over \os(R-\os)} \sum_{\ot}  \FF_R(\ot)
-m^2 {\gv\over\pi} R\sum_{\ot} {\FF_R(\ot) \over \ot(R-\ot)}
\nonumber\\
&&+2m^2 \left({\gv\over\pi}\right)^2
 R\sum_b {\DD_b\over(b-\aa)[(b-\aa)^2-R^2]}
 \sum_{\ot}  \FF_R(\ot)
\lbl{bs-v-0}
\ena

Let us suppose $1<<R<<|\aa|=O(\LL)$, i.e. take the thermodynamical limit
$L\rarr \infty$ with ${R\over L},{\LL\over L}$ fixed, we can
substitute summations with integrals in
the variable $x={\os\over R}$ with ${\gg\over R}\le x\le 1-{1-\gg\over R}$,
where $\gg=\aa-[\aa]$  is the difference between $\aa$ and the nearest
halfinteger $[\aa]$, and evaluate
$$\sum_b {\DD_b\over(b-\aa)[(b-\aa)^2-R^2]}
={\log R\over R^2}+O({1\over R^2})$$
where the interval which decouples yields the non leading contribution
$O({1\over R^2})$.

After performing the previous steps the \BS equation (\rf{bs-v-0}) becomes
\eq
\left({M^2_{\mbox{meson}}\over m^2}-{1\over x(1-x)}\right)\ff(x)=
-{\gv\over\pi}\left[ \int dy~ {\ff(y)\over y(1-y)}+
\left( {1\over x(1-x)}- 2{\gv\over\pi}\log R\right) \int
dy~\ff(y)\right]
\lbl{BS-vett}
\en
It is now easy to check that this equation and its solution are well defined
in the thermodynamical limit. 
In fact when we plug into the previous eq. (\rf{BS-vett}) its solution
\eqa
\ff(x)&=&{\gv\over\pi}{ A +x(1-x) (B- 4A{\gv\over\pi}\log R )
\over 1 - {M^2_{\mbox{meson}}\over m^2} x(1-x)}~~~~
\nonumber\\
&&A=\int_{\gg\over R}^{1-{1-\gg \over R}} dy~\ff(y)~~~~
B=\int_{\gg\over R}^{1-{1-\gg \over R}} dy~ {\ff(y)\over y(1-y)}
\ena
and take the thermodynamical limit, all the logarithmic divergences
cancel and we get a finite wave function and a finite \BS equation
(for a complete discussion of the bound state spectrum see for example
\ct{Ca})).

\sect{The general case}
In the general case the computation is more difficult but it yields
the same kind of results.
We start examining the vacuum energy which can be now written as
\eq
P^-_{\mbox{vacuum}}=N {m^2 L\over 2 \pi} {x\over 1-{\gs\over \pi} x}
{}~~~~
x=\SS(0)
\en
with $x_{min}\le x\le x_{Max}$
We find the same result as with
only the vector current, i.e. eq. (\rf{vacuum}) with the further
constraint
\eq
\left| {\gs\over\pi}\right| \SS(0)<1
\lbl{g-constraint}
\en
in perfect accordance with what we have found in the pure Gross-Neveu
model (\ct{Pe}).

We can now pass to examine the \BS 
equation for the mesonic state (\rf{meson}) which reads
\def\ub{{b-\aa}}
\eqa
{M^2_{\mbox{meson}}\over M^2}\FF_R(\os)&=&
{R^2\over \os(R-\os)} \FF_R(\os)
\nonumber\\
&&+R\sum_{\ot} {\FF_R(\ot) \over \ot}
   \left( {\ju_R\over \os} -{\jd_R\over R-\os} +{\gv\over\pi} \jz_R \right)
\nonumber\\
&&+R\sum_{\ot} {\FF_R(\ot) \over R-\ot}
   \left( {\ju_{-R}\over R-\os} -{\jd_{-R}\over \os} +{\gv\over\pi}
          \jz_{-R} \right)
\nonumber\\
&&+{\gv\over\pi}R  \sum_{\ot} \FF_R(\ot)
 \biggr[-{1\over \os(R-\os)}
       -{\jz_{-R}\over R-\os} +{\jz_{R}\over \os}
\nonumber\\
&&\spazio{+{\gv\over\pi}R  \sum_{\ot} \FF_R(\ot) [}
          +{\gv\over\pi}\biggr(2\sum_\ub {\DD_\ub\over\ub[(\ub)^2-R^2]}
\nonumber\\
&&\spazio{+{\gv\over\pi}R  \sum_{\ot} \FF_R(\ot) [}
          +\jz_{R}\sum_\ub {\DD_\ub\over\ub(\ub-R)}
          +\jz_{-R}\sum_\ub {\DD_\ub\over\ub(\ub+R)}
                      \biggr)
 \biggr]
\nonumber\\
&&-m^2 {\gv\over\pi} R\sum_{\ot} {\FF_R(\ot) \over \ot(R-\ot)}
\ena
In the thermodynamical limit we can again substitute summations with
integrals and compute the leading behaviour of the
different summations involved ($R>0$)
\eqa
\SS(\pm R)=\sum_b {\DD_b\over \ub\pm R}&=& \log \left(R\over |\aa|\right) +O(1)
\nonumber\\
\sum_b {\DD_b\over(\ub)(\ub\pm R)}&=&\mp {\log R\over R}
                                    +O\left({1\over R} \right)
\ena
and
\eqa
\ju_{\pm R}&=& -{G_s(R)+G_p(R)\over 2\pi}
\nonumber\\
\jd_{\pm R}&=& {-G_s(R)+G_p(R)\over
2\pi}
\nonumber\\
\jz_{\pm R}&=& \mp{\log R\over R}(\ju_{\pm R} -\jd_{\pm R})
\ena
with


\eqa
G_s(R)&=&{\pi\over \log {R\over \aa \exp{\pi\over\gs}}}
\nonumber\\
G_p(R)&=&{\pi\over \log {R\over \aa \exp{-{\pi\over\gp}}}}
\lbl{coup-const}
\ena
Here the coupling constants have the same functional dependence on the
momentum as in the usual covariant approach, and hence the same $\bb$
functions  but they depend on the $minus$ component $R^-$ of the 2-momentum
$R^\mu$ instead of depending on its Lorentz invariant modulus $R^\mu R_\mu$.
Notice that if we want to have an asymptotic free theory we have
to set $\gp>0>\gs$.

The final form for the \BS eq. is
\eqa
\left({M^2_{\mbox{meson}}\over M^2}-{1\over x(1-x)}\right)\ff(x)
&=&
\int dy~ {\ff(y)\over y}
   \left[{\ju_R\over x} -{\jd_R\over 1-x}+{\gv\over\pi} R\jz_R\right]
\nonumber\\
&+&
\int dy~ {\ff(y)\over 1-y}
   \left[{\ju_{-R}\over 1-x} -{\jd_{-R}\over x}-{\gv\over\pi} R\jz_{-R}\right]
\nonumber\\
&+&
{\gv\over\pi}\int dy~\ff(y)
  \biggl[-{1\over x(1-x)}-{R\jz_{-R}\over 1-x} +{R \jz_R\over x}
\nonumber\\
&&\spazio{{\gv\over\pi}\int dy~\ff(y)}
  +{\gv\over\pi}\log (R )\left( 2- R \jz_R +R \jz_{-R} \right)
  \biggr]
\nonumber\\
&&
-{\gv\over\pi} \int dy~ {\ff(y)\over y(1-y)}
\lbl{bs-gen}
\nonumber\\
&=& -{G_s(R)\over 2\pi} \left( {1\over x}-{1\over 1-x}\right)
\int dy~ \ff(y)\left({1\over y}- {1\over 1-y}\right)
\nonumber\\
&&+
{G_p(R)\over 2\pi} \left( {1\over x}+{1\over 1-x}-2 {\gv\over \pi}\log R\right)
\int dy~ \ff(y)\left({1\over y} +{1\over 1-y}\right)
\nonumber\\
&&-
{\gv\over \pi} \left( {1\over x}+{1\over 1-x}-2 {\gv\over \pi}\log R\right)
 \left( 1- \log (R) {G_p(R)\over 2\pi} \right)
\int dy~ \ff(y)
\nonumber\\
&&
-{\gv\over\pi} \int dy~ {\ff(y)\over y(1-y)}
\ena
As done in the simplest case of the pure vectorial current interaction
we can again solve this equation and check explicitely that both it
and its solution are finite in the thermodynamical limit.

The equation (\rf{bs-gen}) looks very different from the one obtained
in ref. (\ct{TO}) (eq. 3.23) and one could wonder how it is possible to
recover eq. (3.23) of ref. (\ct{TO}). The key point is that
eq. (\rf{bs-gen}) is finite while the corresponding equation in
(\ct{TO}) is not, therefore we must first drop the subtraction terms,
which make it finite, so ``freeze'' the running coupling constants
to their bare values because they also contribute to the finitness and
endely set the integration interval to $[0,1]$.
Using the previous prescriptions we get from  eq.s (\rf{def0})
$\ju_{\pm R}={\gs-\gp\over 2\pi}, \jd_{\pm R}={\gs+\gp\over2\pi},\jz=0$
and $\log R\equiv 0$ which, when inserted in eq. (\rf{bs-gen}) yield
\eqa
\left({M^2_{\mbox{meson}}\over M^2}-{1\over x(1-x)}\right)\ff(x)
&=&
{\gs\over 2\pi} \left( {1\over x}-{1\over 1-x}\right)
\int_0^1 dy~ \ff(y)\left({1\over y}- {1\over 1-y}\right)
\nonumber\\
&-&
{\gp\over 2\pi} \left( {1\over x}+{1\over 1-x}\right)
\int_0^1 dy~ \ff(y)\left({1\over y} +{1\over 1-y}\right)
\nonumber\\
&-&
{\gv\over\pi}\left( {1\over x(1-x)}\int_0^1 dy~\ff(y)
                   +\int_0^1 dy~ {\ff(y)\over y(1-y)}\right)
\nonumber\\~
\lbl{eq-frozen}
\ena
(for a complete discussion of the bound state spectrum see for example
\ct{Ca})).

\sect{Conclusion.}
We have considered the generalized Gross-Neveu model using DLCQ and
when we examined the vacuum energy, we have found with suprise that
the bare theory has a {\sl non} trivial vacuum, even if this vacuum is very
simple since it is given by the sum of two ``bands''.
Yet the renormalized theory has a trivial vacuum.
It would be nice to see what happens in the infinite momentum frame
(\ct{Ho},\ct{FS}) in order to understand how peculiar of DCLQ these
results are.

We also found  that the running coupling constants emerge in a natural way
(even if there is no trace of running in the ``frozen'' form
eq. (\rf{eq-frozen}) )
but the most interesting result is perhaps the ``universality'' of the
good UV behaviour of the \BS equations in the thermodynamical limit.


\app{}
In this appendix we describe the method used to solve the constraints
and to obtain the $P^-$ generator explicitely.

Our starting point is
to take eq. (\rf{eq-mot-b}) as the quantum constraint, which we multiply
to the left with $\pb^i(y)$ summing over $i$ obtaining
\eqa
i\rd \part_{x-}A(x,y) -m\pb(y)\cdot\pp(x)
-2{\gs+\gp\over N}\pb(y)\cdot\pp(x)~A^*(x,x)
\nonumber\\
-2{\gs-\gp\over N}\pb(y)\cdot\pp(x)~A(x,x)
-4{\gv\over N}\pb(x)\cdot\pp(x)~A(x,y)
=0
\lbl{eq0}
\ena
where we have defined
\eq
A(x,y)=\sum_i\pb^i(y)\cc^i(x)
{}~~~~
A^*(x,y)=\sum_i\cb^i(x)\pp^i(y)
\en
In order to proceed we observe that at the leading order in ${1\over N}$
the operators
\eqa
{1\over N}\pb(x)\cdot\pp(y)
={1\over N}N_\CA[\pb(x)\cdot\pp(y)] +{1\over\rd}\DD(y,x)
\lbl{oper}
\\
\DD(y,x)=\sum_r { e^{i\pi {r\over L} (x-y)} \over 2L} \DD_r ~~~~
\DD_r=\left\{\bear{ll} 1 & r\in\CA \\ 0 & r\in\CC \enar \right.
\lbl{dd}
\ena
commute with everything and hence they can be treated as if they
were classical objects.
This statement deserves a better explanation since it is
fundamental for the explicit solution of the constraints.
The main point is that the operators
$O_{x;y}={1\over N}N_\CA[\pb(x)\cdot\pp(y)]$
enter the expressions of $P^\pm, A$ and when commuting two of such
operators we get
$[O_{x_1;y_1},O_{x_2;y_2}]=
 {1\over N}\left( \usrd \dd_{y_1 x_2} O_{x_1;y_2} +\dd_{y_1 x_2} \DD_{y_2;x_1}
 -(x\lrarr y) \right)$
and this is zero at the leading order in ${1\over N}$.

We can therefore expand $A(x,y)$ in a power series in those operators
(\rf{oper})
\eqa
A(x,y)&=&N~\sum_{n=0}^{\infty} {1\over N^n} A_n(x,y)
\nonumber\\
A_n(x,y)&=&\int_{\{ x_i,y_i\}_{i=1\dots n}}
    A_n(x,y;\{ x_i,y_i\}_{i=1\dots n})~\prod_{i=i}^n
       N_\CA[\pb(x_i)\cdot\pp(y_i)]
\lbl{expansion}
\ena
Then we normal order explicitely the terms of the form
$\pb(y)\cdot\pp(x)$ in eq. (\rf{eq0}),
we use the previous expansion eq. (\rf{expansion}) and we project
onto the different sectors with a different number of operators obtaining
\eqa
&&
\kern-3em
i\rd D_x A_n(x,y)
-\rd(\gs-\gp) \DD(x,y)  A_n(x,x)
-\rd(\gs+\gp) \DD(x,y)  A_n^*(x,x)
=
\nonumber\\
&=&
4\gv N[\pb(x)\cdot\pp(x)] A_{n-1}(x,y)
+2 (\gs-\gp)N[\pb(y)\cdot\pp(x)] A_{n-1}(x,x)
\nonumber\\
&&+2 (\gs+\gp)N[\pb(y)\cdot\pp(x)] A_{n-1}^*(x,x)
-B_n
\lbl{eq1}
\ena
where we have set
\eqa
B_0(x,y)&=&-{m\over\rd} \DD(x,y)
\nonumber\\
B_1(x,y)&=&-{m}N[\pb(y)\cdot\pp(x)]
\nonumber\\
B_n(x,y)&=&0 ~\forall n>1
\ena
and we have defined a new derivative-like operator as
\eq
D_x=\part_x+ 2i\gv \DD(0)
\Llrarr
D^{-1}(x,y)={1\over 2\pi i} \sum_r { e^{i\pi {r\over L} (x-y)} \over
r+\aa}=- (D^{-1}(y,x))^{-1}
\lbl{D}
\en
with
\eq
\aa={\gv\over\pi}~2L\DD(0)={\gv\over\pi}(\LL+\um)
\lbl{def-aa}
\en
where we have used the  C-invariance of the vacuum in order to derive the
equality $2L\DD(0)=\LL+\um$, which is valid for all the test states $|\CA>$
and we have introduced a UV cutoff $\LL\in \interi+\um$ in such a way
that $|r|\le \LL$.

With the help of the previous definitions
we can rewrite eq. (\rf{eq1}) and its hermitian conjugate in an
equivalent form suitable to be solved recursively as



\eq
\CM^{-1}\CA_n=\CC \CA_{n-1} +{} \CN_n
\en
where
\eqa
\CA_n(x,y)&=&\vett{ A_n(x,y)} { A_n^*(x,y)}
\nonumber\\
\CN_n(x,y)&=&{i\over\rd} \vett{ \int_z D^{-1}(x,z) B_n(z,y)}
                  {\int_z D^{-1}(z,x)B _n^*(z,y)}
\ena
and
\eqa
\CM^{-1}(x,y;u,v)
&=&
\mat
{\dd_{x,u}\dd_{y,v}+ i(\gs-\gp)D^{-1}_{x,u} \DD_{u,y} \dd_{u,v} }
{i(\gs+\gp)D^{-1}_{x,u} \DD_{u,y} \dd_{u,v} }
{i(\gs+\gp)D^{-1}_{u,x} \DD_{y,u} \dd_{u,v} }
{\dd_{x,u}\dd_{y,v}+ i(\gs-\gp)D^{-1}_{u,x} \DD_{y,u} \dd_{u,v} }
\nonumber\\~
\lbl{M-1}
\\
\CC(x,y;u,v)
&=&-i\rd
\left(\bear{l}
{2\gv D^{-1}_{x,u} N[\pb_u\cdot\pp_u] \dd_{v,y}
 + (\gs-\gp)D^{-1}_{x,u} N[\pb_y\cdot\pp_u] \dd_{u,v}
}\\
{(\gs+\gp)D^{-1}_{u,x} N[\pb_u\cdot\pp_y] \dd_{u,v}}
\enar\right.
\nonumber\\
&&\left.\bear{l}
{(\gs+\gp)D^{-1}_{x,u} N[\pb_y\cdot\pp_u] \dd_{u,v}}
\\
{2\gv D^{-1}_{u,x} N[\pb_u\cdot\pp_u] \dd_{v,y}
 + (\gs-\gp)D^{-1}_{u,x} N[\pb_u\cdot\pp_y] \dd_{u,v}
}
\enar\right)
\nonumber\\~~
\ena

\app{}
In this appendix we give some formulae which can be useful in checking
the computations.

For computing the actual expression of $P^-$ eq. (\rf{P-})
we have used the fact that we can write
\eqa
P^-&=&{N~m\over 2} \sum_{n=0}^\infty {1\over N^n}
  \int_x A_{(n)}(x,x)+A_{(n)}^*(x,x)
={N~m\over 2} \sum{1\over N^n} \vet{\dd_{x y}}{\dd_{x y}} \CA_{(n)}
\nonumber\\
&=&{N~m\over 2}\sum{1\over N^n} \vec{1} \CA_{(n)}
\ena
and we give

The Fourier expansion that we use is given by:
\eq
X_{x_1\dots x_n}=\sum_{r_1\dots r_n}
{ e^{{i\pi\over L} \left( r_1 x_1+\dots+ r_n x_n\right)}
 \over(2L)^{n\over 2} }
X_{r_1\dots r_n}
\en

In the following we give the Fourier components of matrices as
$X_{a,b;-r,-s}$, of vectors as $Y_{a,b}$ and of transposed vectors as
$Y_{-r,-s}$ because it is hence immediate to multiply for instance two
matrices $X,Y$ using the realtion
\eq
(X Y)_{a,b;-r,-s}= \sum_{p,q} X_{a,b;-p,-q} Y_{p,q;-r,-s}
\en

We can write the expression for the matrix $\CM$ defined in
eq. (\rf{M-1}) as well as sketch its derivation.
\eqa
\CM_{x,y;u,v}
&=&
\uno
+\CM_{(0)x,y;u,v}
\nonumber\\
\CM_{(0)a,b;-r,-s}
&=&\mat{Q_{b,-r-s}}{-F_{b,-r-s}}
     {-F_{-b,r+s}^*}{Q_{-b,r+s}^*}
 \dd_{a+b,r+s}
\nonumber\\
Q_{b,n}
&=&
{\DD_b\over b-\aa+n}
{  {\gs-\gp\over 2\pi} + {\gs\over\pi} {\gp\over\pi} \SS(n)
    \over 1 - {\gs-\gp\over 2\pi} (\SS(n)+\SS(-n))
            - {\gs\over\pi} {\gp\over\pi}\SS(n)\SS(-n) }
\nonumber\\
F_{b,n}&=&
{\DD_b\over b-\aa+n}
{ - {\gs+\gp\over 2\pi}
    \over 1 - {\gs-\gp\over 2\pi} (\SS(n)+\SS(-n))
            - {\gs\over\pi} {\gp\over\pi}\SS(n)\SS(-n) }
\ena
The previous formulae can be derived if we start writing
\eq
\CM^{-1}= \uno +\CI ~~~~
\CI_{x,y;u,v}=I_{x,y;u} \dd_{u v}
\en
and then we use the obvious formula
\eq
\CM=\uno+\sum_{n=1}^\infty (-)^n \CI^n
\en
where 
\eq
\CI^n_{x,y;u,v}=\int_z I_{x,y;z} {\bar I}^{n-1}_{z;u}~\dd_{uv} ~~~
{\bar I}_{x;u}\equiv I_{x,x;u}
\en

Other useful formulae are
\eq
\CN_{(0)a,b}={L\over 2\pi }\vett{ { N_{(0)b}} }{ { N_{(0)-b}}^* }
    \dd_{a+b,0}~~~~
{ N_{(0)b}}
=m {\DD_b \over b-\aa}
\en

\eq
\CN_{(1)a,b}
={L\over 2\pi }\vett{ { N_{(1)a,b}} }{ { N_{(1)-a,-b}}^* } ~~~~
{ N_{(1)a,b}}
=-{1\over a+\aa}
N[\pb_{-b}\cdot\pp_{a}]
\en

and the intermediate results
\eqa
(\CC\CM\CN_{(0)}+\CN_{(1)})_{a,b}
&=&{L\over 2\pi }\vett{ { S_{a,b}} }{ { S_{-a,-b}}^* }
\nonumber\\
S_{a,b}
&=&
-{m\over 1-{\gs\over\pi}\SS(0)}{1\over a+\aa}
\left( N[\pb_{-b}\cdot\pp_a]
     + {\gv\over\pi}{\DD_b\over b-\aa}\sum_p N[\pb_{p}\cdot\pp_{p+a+b}]
\right)
\nonumber\\~
\ena

\eqa
(\vec{1} \CM\CC\CM)_{-r,-s}
&=&\vet{S_{(1),-r,-s}}{S_{(1),r,s}^*}
\nonumber\\
S_{(1),-r,-s}
&=&
{1\over 1-{\gs\over\pi}\SS(0)}\Biggl[
 {\gs\over\pi}\left( {1\over -s+\aa} +\jz_{r+s}\right)
  \sum_p N[\pb_{p}\cdot\pp_{p-r-s}]
\nonumber\\
&&
+\ju_{r+s}   \sum_p {N[\pb_{p}\cdot\pp_{p-r-s}]\over p+\aa}
+\jd_{r+s}   \sum_p {N[\pb_{p+r+s}\cdot\pp_{p}]\over p+\aa}
\Biggr]
\nonumber\\~
\ena
where the symbols $\jz,\ju,\jd$ are defined in eq. (\rf{def0}).

\end{document}